\documentclass[prl,twocolumn,a4paper,aps,superscriptaddress]{revtex4}
\usepackage{graphicx}
\usepackage{dcolumn}
\usepackage{bm}
\usepackage{amssymb}
\usepackage{epsfig}
\begin{document}
\title{SQUIPT - Superconducting Quantum Interference Proximity Transistor}
\author{F. Giazotto}
\email{f.giazotto@sns.it}
\affiliation{NEST CNR-INFM and Scuola Normale Superiore, Piazza dei
Cavalieri 7, I-56126 Pisa, Italy}
\author{J. T. Peltonen}
\affiliation{Low Temperature Laboratory, Helsinki University of Technology, P.O. Box 3500, 02015 TKK, Finland}
\author{M. Meschke}
\affiliation{Low Temperature Laboratory, Helsinki University of Technology, P.O. Box 3500, 02015 TKK, Finland}
\author{J. P. Pekola}
\affiliation{Low Temperature Laboratory, Helsinki University of Technology, P.O. Box 3500, 02015 TKK, Finland}
\date{\today}
\begin{abstract}
We present the realization and characterization of a novel-concept interferometer, the superconducting quantum interference proximity transistor (SQUIPT). Its operation relies on the modulation with the magnetic field of the density of states of a proximized metallic wire embedded in a superconducting ring.
Flux sensitivities down to $\sim 10^{-5}\,\Phi_0$Hz$^{-1/2}$ can be achieved even for a non-optimized design, with an intrinsic dissipation ($\sim 100$ fW) which is several orders of magnitude smaller than in conventional superconducting interferometers.
Our results are in agreement with the theoretical prediction of the SQUIPT behavior, and suggest that optimization of the device parameters would lead to a large enhancement of sensitivity for the detection of tiny magnetic fields.
The features of this setup and their potential relevance for applications are further discussed.
\end{abstract}
\maketitle

\emph{Proximity effect} \cite{degennes} is a phenomenon which can be described as the induction of superconducting correlations into a normal-type conductor \cite{heersche,jherrero,keizer,morpurgo,kasumov,cleuziou,doh,xiang, pothier,courtois,morpurgo1,giazotto,baselmans}.
One striking consequence of this effect is the modification of the local density of states (DOS) in the normal metal \cite{belzig,belzig2,gueron,sueur}, and the opening of a minigap \cite{belzig2,belzig,zhou} whose amplitude can be controlled by changing the macroscopic phase of the superconducting order parameter \cite{belzig,sueur}.
Here we report the realization of a novel interferometer, the superconducting quantum interference proximity transistor (SQUIPT), whose operation relies on the modulation with the magnetic field of the DOS of a proximized metal embedded in a superconducting loop.
Flux sensitivities down to $\sim 10^{-5}\,\Phi_0$Hz$^{-1/2}$ ($\Phi_0 \simeq 2\times 10^{-15}$ Wb is the flux quantum) can be achieved even for a non-optimized design, with an intrinsic dissipation which is several orders of magnitude smaller than in conventional superconducting interferometers \cite{clarke,tinkham,likharev}.
Optimizing the device parameters promises to largely increase the sensitivity for the detection of tiny magnetic fields.

\begin{figure}[t!]
\includegraphics[width=8.3cm]{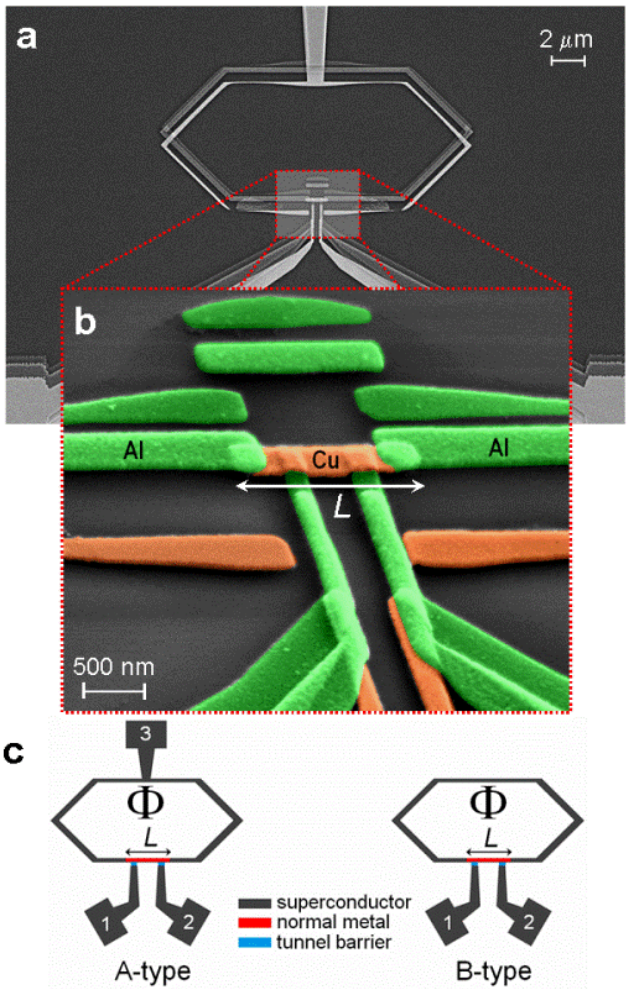}
\caption{\label{fig1} The SQUIPT. (a) Scanning electron micrograph of a typical superconducting quantum interference proximity transistor (SQUIPT). An aluminum (Al) superconducting ring is interrupted by a normal metal copper (Cu) wire in good electric contact with it. Two Al tunnel junctions (bottom of the image) are connected to the normal metal to allow the device operation.
(b) A pseudo-color blow-up of the junction region showing the Al/Cu/Al superconductor-normal metal-superconductor proximity junction as well as the two Al electrodes (of width $\sim 200$ nm) connected via an AlOx tunnel barrier to the Cu wire.
The latter is overlapped laterally by the Al leads for about 250 nm.
The structure replicas due to the shadow-mask fabrication procedure are visible.
(c) Sketch of the two device geometries investigated.
The superconducting ring extends into an additional third lead in the A-type configuration. The electrodes labeled 1, 2 and 3 are used to operate the SQUIPT.
$L\simeq 1.5$ $\mu$m denotes the total length of the normal metal region, while $\Phi$ is the applied magnetic flux threading the loop.
}
\end{figure}
One typical SQUIPT fabricated with electron-beam lithography is shown in Fig. 1(a). It consists of an aluminum (Al) superconducting loop interrupted by a copper (Cu) normal metal wire in good electric contact with it. Furthermore, two Al electrodes are tunnel-coupled to the normal region to allow the device operation.
A blow-up of the sample core [see Fig. 1(b)] displays the Cu region of length $L\simeq 1.5\,\mu$m and width $\simeq 240$ nm coupled to the tunnel probes and the superconducting loop. The SQUIPTs were implemented into two different designs [see Fig. 1(c)], namely, the A-type configuration, where the loop extends into an additional third lead, and the B-type configuration which only contains two tunnel probes.
The ring geometry allows to change the phase difference across the normal metal-superconductor boundaries through the application of an external magnetic field which gives rise to a total flux $\Phi$ through the loop area.
This modifies the DOS in the normal metal, and hence the transport through the tunnel junctions.

\begin{figure}[t!]
\includegraphics[width=\columnwidth]{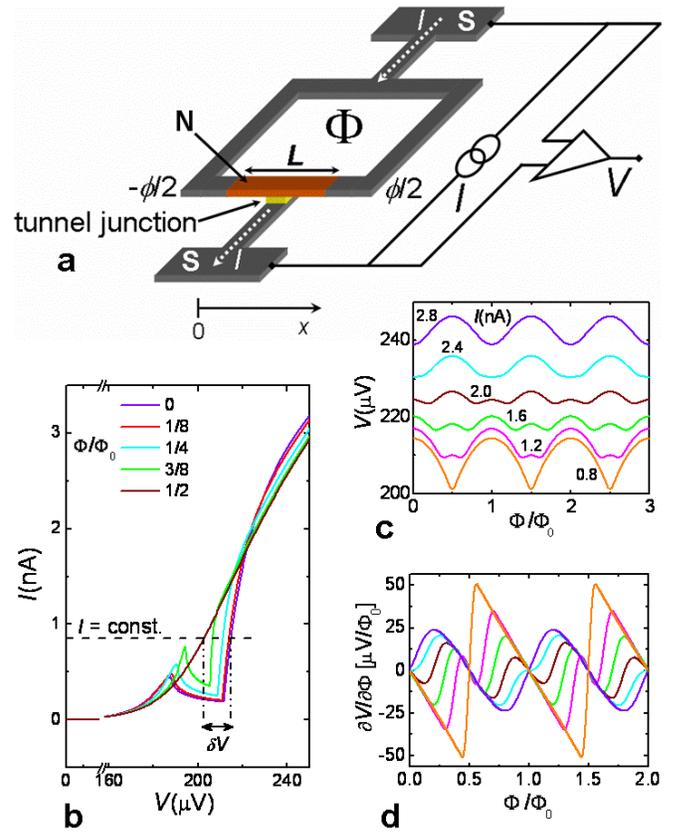}
\caption{\label{fig2} SQUIPT configuration and its predicted behavior. (a) Schematic drawing of the SQUIPT implemented in the A-type configuration. The superconducting (S) tunnel probe with normal-state resistance $R_T$ is placed in the middle of the normal metal (N) wire, i.e, at $x=0$. An electric current $I$ is fed into the circuit, while the voltage drop $V$ is recorded as a function of the applied magnetic flux $\Phi$.
$\phi$ is the macroscopic quantum phase of the superconducting order parameter.
(b) SQUIPT quasiparticle current-voltage characteristics calculated for some values of the applied flux $\Phi$.
When biased with a constant current $I$ a voltage modulation $V(\Phi)$ with amplitude $\delta V$ appears across the device as a function of the applied magnetic flux.
(c) Voltage modulation $V(\Phi)$ for some values of the bias current.
(d) Flux-to-voltage transfer function $\partial V/\partial \Phi$ calculated for the same current values as in panel (c).
All calculations where performed at temperature $T=0.1 T_c$ by setting $E_{Th}=4\,\mu$eV which is similar to that in our devices. Here, $T_c\simeq 1.3$ K is the superconducting critical temperature,
$E_{Th}=\hbar D/L^2$ is the Thouless energy, i.e., the characteristic energy scale of the normal metal region,
$D=110$ cm$^2$/s is the electron diffusion constant in Cu, and $\hbar$ is the Planck's constant. Furthermore, we set $\Delta_0=200\,\mu$eV as the zero-temperature Al superconducting gap, and $R_T=50$ k$\Omega$.
}
\end{figure}

Insight into the interferometric nature of the SQUIPT can be gained by analyzing first of all the theoretical prediction of its behavior.
Figure 2(a)
 sketches the simplest implementation of the device in the A-type configuration, i.e., that with just one junction tunnel-coupled to normal metal.
For simplicity we suppose the tunnel probe (with resistance $R_T$) to be placed in the middle of the wire, and to feed a constant electric current $I$ through the circuit while the voltage drop $V$ is recorded as a function of $\Phi$.
In the limit that the kinetic inductance of the superconducting loop is negligible, the magnetic flux fixes a phase difference $\phi=2\pi \Phi/\Phi_0$ across the normal metal, where $\Phi_0=\pi \hbar/e$ is the flux quantum, $\hbar$ the Planck's constant, and $e$ the electron charge.
Figure 2(b) shows the low-temperature quasiparticle current-voltage ($I-V$) characteristic of the SQUIPT calculated at a few selected values of $\Phi$.
The calculations were performed for parameters similar to those of our structures \cite{sns}.
It clearly appears that while for $\Phi=0$, i.e., when the minigap in the normal metal is maximized \cite{sueur,zhou}, the characteristic resembles that of a superconductor-insulator-superconductor junction \cite{tinkham}, for $\Phi=\Phi_0/2$ the characteristic corresponds to that of a normal metal-insulator-superconductor contact, with the minigap suppressed \cite{sueur,zhou}.
The SQUIPT thus behaves as a flux-to-voltage transformer whose response $V(\Phi)$ (and amplitude $\delta V$) depends on the bias current $I$ through the tunnel junction. The interferometer voltage modulation $V(\Phi)$ is shown in Fig. 2(c) for different values of $I$.
In particular, $V(\Phi)$ is strongly dependent on the bias current, the latter determining the exact shape of the device response.
Note the change of concavity of $V(\Phi)$ which occurs as the bias current exceeds the point where the $I-V$ characteristics cross.
One relevant figure of merit of the SQUIPT is represented by the flux-to-voltage transfer function, $\partial V/\partial \Phi$, which is shown in Fig. 2(d).
It turns out that $\partial V/\partial \Phi$ is a non-monotonic function of the bias current, as well as its sign depends on the specific value of $I$.

\begin{figure}[t!]
\includegraphics[width=\columnwidth]{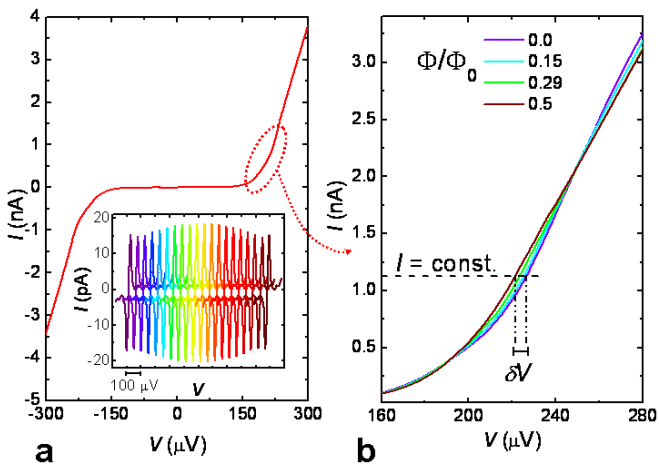}
\caption{\label{fig3} Magnetic-field dependence of the current-voltage characteristic. 
(a) Current-voltage characteristic of an A-type SQUIPT with normal-state resistance $R_T\simeq 50$ k$\Omega$ measured at $68$ mK.
The inset shows the enlargement around zero bias of the characteristic of the same device measured at $53$ mK for several applied magnetic field values.
The peak appearing around zero-bias, and with magnitude $I\simeq 17$ pA, is the Josephson current flowing through the structure, and its magnitude shows a periodic modulation over a magnetic flux quantum $\Phi_0$. The curves are horizontally offset for clarity, and correspond to a magnetic field intensity increasing with a step of $\sim 10^{-2}$ Oe. For the present structure $\Phi_0$ corresponds to an applied field $B=\Phi_0/A\simeq 0.17$ Oe, where $A\simeq 120$ $\mu$m$^2$ is the loop area.
(b) Blow-up of the current-voltage characteristic of the same device for larger bias voltages measured at $53$ mK for some values of the applied magnetic flux $\Phi$ up to $\Phi_0/2$.
Also shown is the voltage modulation amplitude $\delta V$ occurring when biasing the SQUIPT with a constant current $I$.
All measurements on the A-type structure were performed through electrodes 1 and 3 [see Fig. 1(c)].
}
\label{fig3}
\end{figure}
Figure 3(a) displays the experimental low-temperature $I-V$ characteristic of a device implemented in the A-type configuration. The curve resembles that of a typical superconducting tunnel junction, where the onset of large quasiparticle current is set by the energy gap ($\Delta_0\simeq 200\,\mu$eV in our samples).
The absence of the peak
in the experimental curves, as compared to the theoretical ones in Fig. 2(b), could originate from broadening
due to inelastic scattering or finite quasiparticle lifetime
in the superconductor \cite{timofeev}.
A deeper inspection reveals, however, that the characteristic is modulated by the presence of an applied magnetic field.
The effect is clearly visible in Fig. 3(b) which shows the blow-up of the curve at large bias voltage for some values of the applied flux up to $\Phi_0/2$.
Such a modulation is of coherent nature, and stems from magnetic field-induced control of the DOS in the normal metal.
In addition to the quasiparticle current, a Josephson coupling is observed at the lowest temperatures, and manifests itself as a peak around zero bias in the $I-V$ characteristic [see the inset of Fig. 3(a)].
The supercurrent, which
is expected to exist in proximized structures like the present one \cite{belzig}, obtains values as high as $\simeq 17$ pA at 53 mK. It is modulated by the applied flux with the same periodicity as for the quasiparticle current.

\begin{figure}[t!]
\includegraphics[width=\columnwidth]{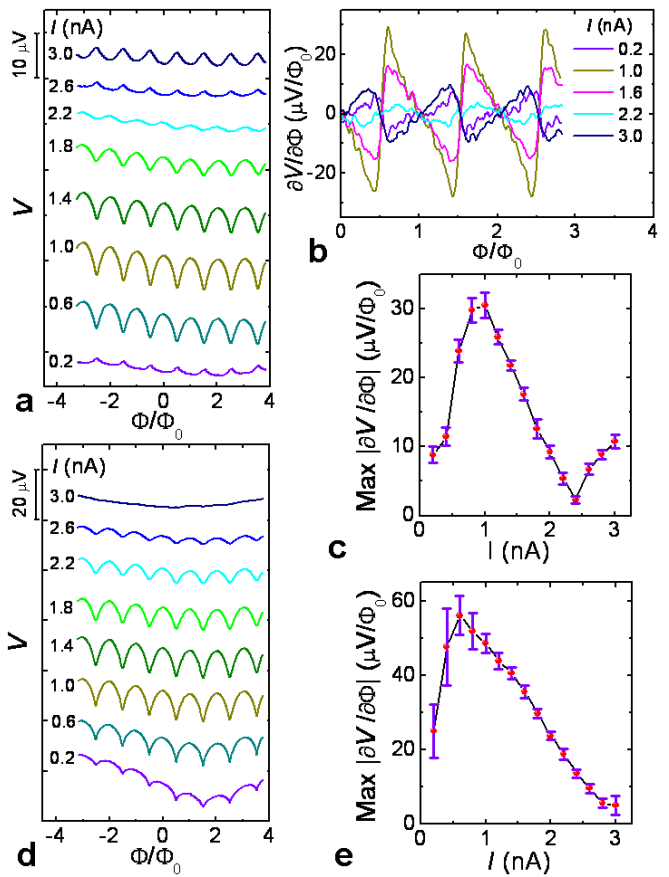}
\caption{ Magnetic-field dependence of the voltage modulation and flux-to-voltage transfer function. 
(a) Voltage modulation $V(\Phi)$ of an A-type structure measured at 54 mK for several values of the bias current $I$. 
For the present device $\delta V$ obtains values as large as $\sim 7\,\mu$V at 1 nA. 
Note the change of concavity of $V(\Phi)$ for different values of $I$.
The curves are vertically offset for clarity.
(b) Flux-to-voltage transfer function $\partial V/\partial \Phi$ of the same A-type structure at 54 mK at selected bias currents.
(c) Maximum value of $|\partial V/\partial \Phi|$ versus injection current for A-type structure at $54$ mK. 
(d) Voltage modulation $V(\Phi)$ of a B-type structure measured at 53 mK for several values of the bias current $I$. 
For this device the maximum amplitude of the voltage modulation is $\delta V \sim 12\,\mu$V at 1 nA. 
The curves are vertically offset for clarity.
(e) Maximum value of $|\partial V/\partial \Phi|$ versus injection current for the same B-type structure at $53$ mK.
Measurements on the B-type structure were performed through electrodes 1 and 2 [see Fig. 1(c)].
}
\label{fig4}
\end{figure}

The full $V(\Phi)$ dependence for the same A-type sample at several values of the bias current is shown in Fig. 4(a).
As expected [see Fig. 2(c)], the modulation amplitude $\delta V$ is a non-monotonic function of $I$, while $V(\Phi)$ shows changing of concavity whenever the bias current exceeds the crossing points of the current-voltage characteristic [see Fig. 3(b)]. 
In this sample $\delta V$ obtains values as large as $\sim 7\,\mu$V at 1 nA. 
The corresponding transfer function is displayed in Fig. 4(b) for a few bias currents.
Resemblance with theoretical prediction of Fig. 2(c) and 2d is obvious. 
In such a case $|\partial V/\partial \Phi|$ as large as $\simeq 30\,\mu\text{V}/\Phi_0$ is obtained at 1 nA. 
The maximum of $|\partial V/\partial \Phi|$ for the same SQUIPT is displayed in Fig. 4(c), and highlights the expected non-monotonic dependence on $I$.

Figure 4(d) and 4e show $V(\Phi)$ and the maximum of $|\partial V/\partial \Phi|$, respectively, for a B-type SQUIPT .
$\delta V$ obtains in this case values as high as $\sim 12\,\mu$V at 1 nA, while $|\partial V/\partial \Phi|$ is maximized at 0.6 nA where it reaches $\simeq 60\,\mu\text{V}/\Phi_0$.
We emphasize that these values are larger by almost a factor of two than those obtained in the A-type device. This is to be expected since in a B-type sample $V(\Phi)$ is probed across two tunnel junctions in series. This doubles the SQUIPT response.
The above results are roughly $\sim 50-60\%$ of those predicted by our calculations [see Fig. (2)], which can be ascribed either to the uncertainty in the precise determination of the device parameters \cite{sns} or to non-ideal phase biasing of the interferometers \cite{sueur}.
In the limit of negligible geometric inductance of the loop ($\sim 40$ pH for our rings), a phase difference $\phi=2\pi \Phi/\Phi_0$ can be induced across the normal metal if the phase accumulated in the superconductor is much smaller than that accumulated in the wire, i.e., if the
ratio between their respective kinetic inductances is much smaller than unity.
The correction factor to the actual phase bias determined by such a ratio (estimated to be around $0.3-0.5$ for our devices) may prevent the full closing of the minigap, thus weakening the SQUIPT response.

\begin{figure}[t!]
\includegraphics[width=\columnwidth]{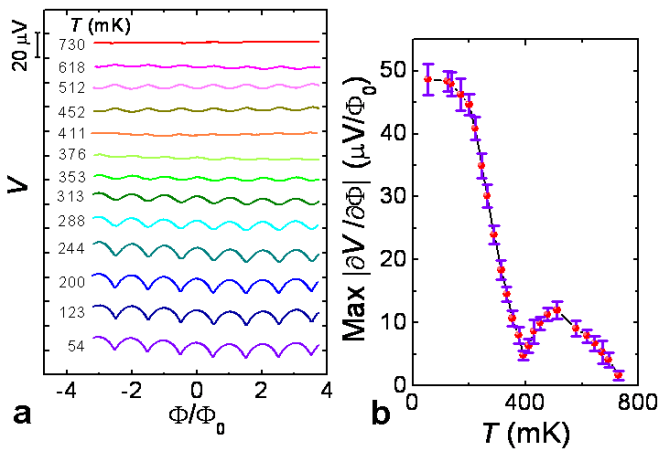}
\caption{ Temperature dependence of the voltage modulation and flux-to-voltage transfer function.
(a) Voltage modulation $V(\Phi)$ of the B-type structure measured at several bath temperatures for $I=1$ nA.
$V(\Phi)$ changes concavity by increasing the temperature between 376 mK and 411 mK.
The curves are vertically offset for clarity.
(b) Maximum value of $|\partial V/\partial \Phi|$ versus temperature for the same device at $I=1$ nA. This behavior resembles in part that of Fig. 4(c) observed for different bias current  values.
}
\label{fig5}
\end{figure}

The role of temperature ($T$) is shown in Fig. 5(a) which displays $V(\Phi)$ at 1 nA for several increasing temperatures for a B-type sample. $\delta V$ initially monotonically decreases with increasing $T$ up to $\sim 400$ mK, then it starts to increase again and it is almost suppressed at $730$ mK.
The full temperature dependence of the maximum of $|\partial V/\partial \Phi|$ at 1 nA is shown in Fig. 5(b) for the same device, and reflects the above non-monotonic behavior similarly to that observed for different bias currents [see Fig. 4(c)].

Compared to conventional DC superconducting quantum interference devices (SQUIDs) \cite{clarke,tinkham,likharev}, power dissipation ($P$) is dramatically suppressed in the SQUIPT.
In our devices we have $P\sim 10^2$ fW, which can be further reduced by simply increasing the resistance of the probing junctions.
This power is 4 - 5 orders of magnitude smaller than that in conventional DC SQUIDs, which makes the SQUIPT ideal for applications where very low dissipation is required.

We shall finally comment on another figure of merit of the SQUIPT, namely, its noise-equivalent flux (NEF) or ``flux sensitivity'' defined as $\text{NEF}=\left\langle V_N^2\right\rangle^{1/2}/|\partial V/\partial \Phi|\delta\nu^{1/2}$ \cite{likharev}, where $V_N$ is the voltage noise of the interferometer within the frequency band $\delta \nu$.
In our experiment we can provide an upper estimate for NEF, since it is believed to be limited mainly by the preamplifier noise. With a typical rms noise of $\sim 1.2$ nV/$\sqrt{\text{Hz}}$ in our setup, we estimate $\text{NEF}\simeq 2\times10^{-5}\Phi_0/\sqrt{\text{Hz}}$ at best, which should be substantially higher than the SQUIPT intrinsic NEF. We note that the preamplifier contribution to the noise can be made negligible by increasing $|\partial V/\partial \Phi|$, i.e., by optimizing the SQUIPT parameters
 and the phase bias as well.
Our calculations show that by replacing niobium (Nb) as the superconductor (with $\Delta_0\simeq 1.5$ meV), and by shortening $L$ down to $150$ nm, $|\partial V/\partial \Phi|$ as large as $\sim 2.5$ mV$/\Phi_0$ could be achieved yielding $\text{NEF}\simeq 4\times 10^{-7}\Phi_0/\sqrt{\text{Hz}}$. The device intrinsic noise deserves however further investigation.

The SQUIPT has a number of features which make it attractive for a variety of applications:
(1) only a simple DC read-out scheme is required, similarly to DC SQUIDs;
(2) either current- or voltage-biased measurement can be conceived depending on the setup requirements;
(3) a large flexibility in the fabrication parameters and materials, such as semiconductors \cite{doh,xiang,morpurgo1,giazotto}, carbon nanotubes \cite{jherrero,morpurgo,cleuziou} or graphene \cite{heersche} instead of normal metals, is allowed to optimize the response and the operating temperature (to this end superconducting V \cite{garcia} or Nb are suitable candidates);
(4) ultralow dissipation ($\sim1\ldots100$ fW) which makes it ideal for nanoscale applications;
(5) ease of implementation in a series or parallel array (depending on the biasing mode) for enhanced output;
(6) ease of integration with superconducting refrigerators \cite{rmp} to actively tune the device working temperature.
Our approach opens the way to magnetic-field detection based on ``hybrid'' interferometers
which take advantage of the flexibility intrinsic to proximity metals.

We gratefully acknowledge O. Astafiev, L. Faoro, R. Fazio, M. E. Gershenson, T. T. Heikkil$\ddot{\text{a}}$, L. B. Ioffe, V. Piazza, P. Pingue, F. Portier, H. Pothier, H. Rabani, F. Taddei, and A. S. Vasenko for fruitful discussions. The work was partially supported by the INFM-CNR Seed project ``Quantum-Dot Refrigeration: Accessing the $\mu$K Regime in Solid-State Nanosystems'', and by the NanoSciERA project ``NanoFridge''.

\section{Supplementary information, methods used in sample fabrication, measurements and theoretical analysis}

\textit{Fabrication details and experimental setup.}
The samples were fabricated at the Low Temperature Laboratory at Helsinki University of Technology by standard three-angle shadow-mask evaporation of the metals through a conventional suspended resist mask in a single vacuum cycle.
Initially, a bi-layer PMMA/copolymer resist was spun on an oxidized Si wafer, onto which the structures were patterned using electron-beam lithography.
In the electron-gun evaporator, the chip was first tilted to an angle of $25^{\circ}$ with respect to the source, and approximately 27 nm of Al was evaporated to create the superconducting electrodes of the probe tunnel junctions. To form the tunnel barriers, the sample was exposed to 4.4 mbar of oxygen for five minutes, and consequently tilted to $-25^{\circ}$ for the deposition of 27 nm of Cu forming the normal metal island. Immediately after this, the chip was tilted to the final angle of $9^{\circ}$, and 60 nm of Al was deposited to form the superconducting loop and the transparent normal metal-superconductor contacts.
The magneto-electric characterization was performed at NEST CNR-INFM by cooling the devices with a filtered $^3$He/$^4$He dilution refrigerator down to $\simeq 50$ mK. Current and voltage were measured with room-temperature preamplifiers.

\textit{Theoretical model.}
The considered system consists of a one-dimensional diffusive normal-metal wire of length $L$ in good electric contact with two superconducting leads which define a ring [see Fig. 2(a)].
The contact with the superconductors allows superconducting correlations to be induced into the normal-metal region through proximity effect \cite{degennes}, that is responsible for the modification of the density of states (DOS) in the wire \cite{belzig2}, as well as for the Josephson current to flow through the superconductor-normal metal-superconductor structure \cite{belzig}.
The proximity effect in the normal-metal region of the SQUIPT can be described with the Usadel equations \cite{usadel}
which can be written as \cite{usadel,belzig}
\begin{eqnarray}
\hbar D\partial^2_x\theta =-2iE\sinh(\theta) +\frac{\hbar D}{2}
\left(\partial_x\chi \right)^2\sinh(2\theta) \nonumber\\
\text{sinh}(2\theta)\partial_x \theta\partial_x \chi+\text{sinh}^2(\theta)\partial_x^2\chi=0,
\label{usadeleq}
\end{eqnarray}
where
$D$ is the diffusion constant and $E$ is the energy relative to the chemical potential in the superconductors.
$\theta$ and $\chi$ are, in general, complex scalar functions of position $x$ and energy.
For perfectly-transmitting interfaces the boundary conditions at the normal metal-superconductor contacts [i.e., $x=\pm L/2$, see Fig. 2(a)] reduce to $\theta(\pm L/2)=\text{arctanh}(\Delta/E)$ and $\chi(\pm L/2)=\pm\phi/2$,
where $\phi$ is the phase difference across the normal metal-superconductor boundaries, and $\Delta$ is the superconducting order parameter.
For simplicity we chose a step-function form for the order parameter, i.e., constant in the superconductor and zero in the normal-metal region, although $\Delta$ is in principle position-dependent and can be determined self-consistently \cite{belzig}.
The DOS in the normal-metal region normalized to the DOS at the Fermi level in the absence of proximity effect is given by $N_N(x,E,\phi)= \mbox{Re}\left\{\cosh\left[\theta(x,E,\phi)\right]\right\}$.
From the numerical solution of Eqs. (\ref{usadeleq})
we get the DOS as a function of position and energy for any given $\phi$.
In particular, the DOS is an even function of energy, and exhibits a minigap ($E_g$) for $|E|\leq E_g$ \cite{belzig2} whose magnitude depends in general on the Thouless energy $E_{Th}=\hbar D/L^2$, the characteristic energy scale for the normal region, and on $\phi$.
The minigap $E_g$ is maximum for $\phi=0$ and decreases with increasing $\phi$, vanishing at $\phi=\pi$ \cite{zhou} (the behavior is $2\pi$-periodic in $\phi$).

The quasiparticle current ($I_{qp}$) through the superconducting tunnel junction biased at voltage $V$ [see Fig. 2(a)] can be calculated from \cite{wolf}
\begin{equation}
I_{qp}=\frac{1}{eR_T}\int dE N_S(E-eV)N_N(E)[f_0(E-eV)-f_0(E)],
\end{equation}
where $e$ is the electron charge, $R_T$ is the tunnel junction resistance, $N_S(E)=|E|/\sqrt{E^2-\Delta^2}\Theta (E^2-\Delta^2)$ is the Bardeen-Cooper-Schrieffer normalized DOS in the superconductor, $\Theta (z)$ is the Heaviside step function, $f_0(E)=(1+\exp[E/(k_{\text{B}}T)])^{-1}$ is the Fermi-Dirac distribution function at temperature $T$, and $k_B$ is the Boltzmann constant.
For any bias current ($I_{bias}$) imposed across the SQUIPT, the voltage response $V(\phi)$ is determined from the solution of the equation $I_{bias}-I_{qp}(x,\phi,T)=0$.
From this, we calculate the flux-to-voltage transfer function $\partial V/\partial \phi$.




\end{document}